\documentclass[a4paper]{article}

\usepackage{icrc2013}
\newcommand{\teq}{t_{\rm EQ}}

\title{Tracing the Interplay between Non-Thermal Dark Matter 
and Right-Handed Dirac Neutrinos with LHC Data}

\shorttitle{Tracing the Interplay between non-thermal DM and Dirac
  $\nu_R$ with LHC data}

\authors{
Luis A. Anchordoqui$^{1}$,
Haim Goldberg$^{2}$,
Brian Vlcek$^{1}$
}

\afiliations{
$^1$ Department of Physics, 
University of Wisconsin-Milwaukee,
 Milwaukee, WI 53201, USA\\
$^2$ Department of Physics,
Northeastern University, Boston, MA 02115, USA \\
}

\email{brian.vlcek@gmail.com}

\abstract{Llight-element abundances probing big bang nucleosynthesis
  (BBN) and precision data from cosmology probing the cosmic microwave
  background (CMB) decoupling epoch have hinted at the presence of
  extra relativistic degrees of freedom. This is widely referred to as
  ``dark radiation'', suggesting the need for new light states in the
  UV completion of the standard model (SM). We provide a brief and
  concise overview of the current observational status of such dark
  radiation and investigate the interplay between two possible
  interpretations of the extra light states: the right-handed partners
  of three Dirac neutrinos (which interact with all fermions through
  the exchange of a new heavy vector meson $Z'$) and dark matter (DM) particles
  that were produced through a non-thermal mechanism, such us late
  time decays of massive relics.  This model ties together
  cosmological indications of extra light states and the production of
  the heavy vector particle at the Large Hadron Collider (LHC).}

\begin{document}
\maketitle

\begin{figure}[ht]
  \centering
  \includegraphics[width=0.45\textwidth]{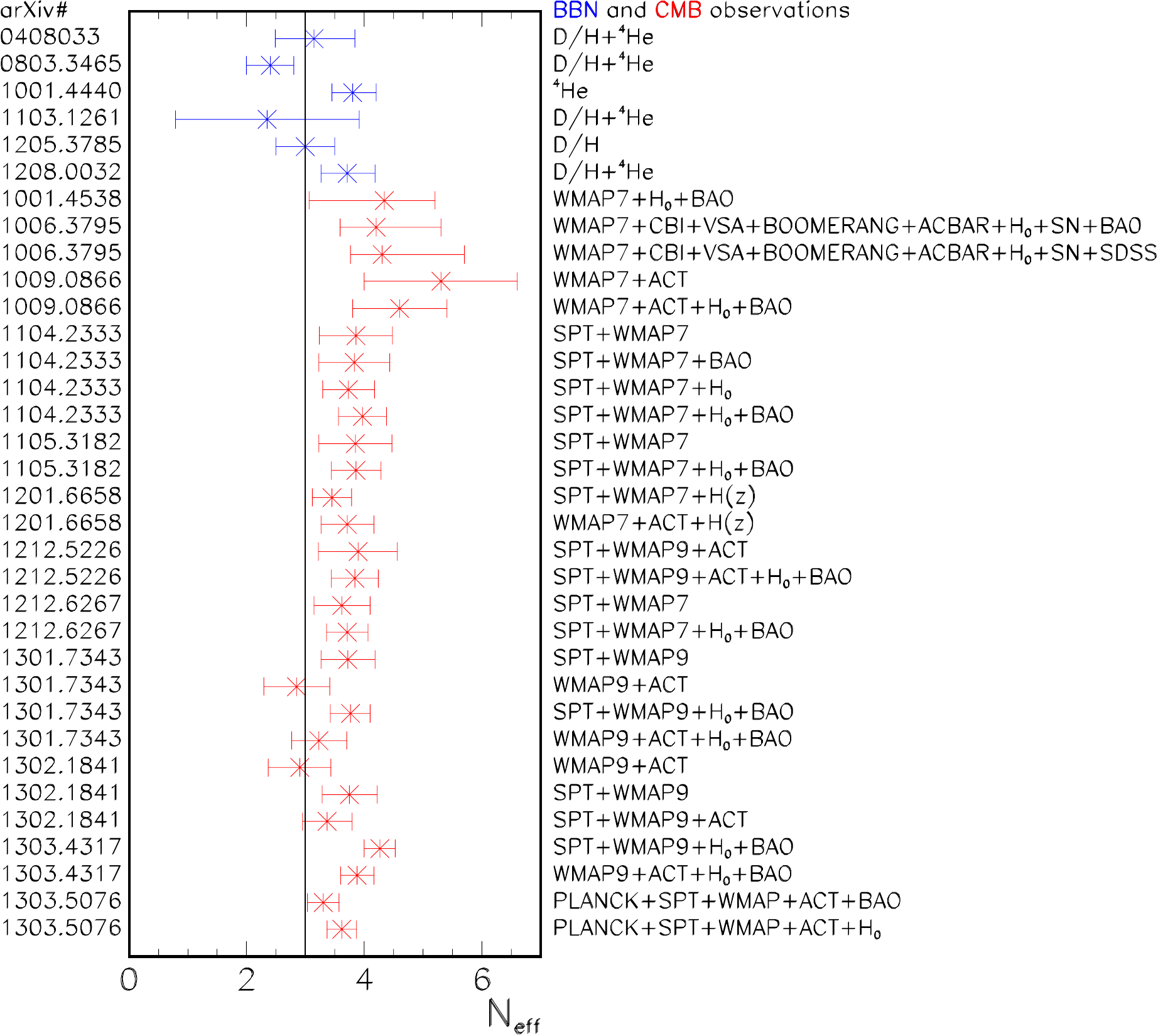}
\caption{A selection of the most recent cosmological $N_{\rm eff}$
  measurements and the $1\sigma$
  confidence  intervals from  various combinations of models and data
  sets. BBN findings are shown in blue and those from the CMB
  epoch in red.}   
  \label{fig:uno}
 \end{figure}

To accommodate new physics in the form of extra relativistic degrees
of freedom (r.d.o.f.)  it is convenient to account for the extra
contribution to the SM energy density, by normalizing it to that of an
``equivalent'' neutrino species. The number of ``equivalent'' light
neutrino species, $N_{\rm eff} \equiv (\rho_{\rm R} -
\rho_\gamma)/\rho_{\nu_L}$, quantifies the total ``dark'' relativistic
energy density (including the three left-handed SM neutrinos) in units
of the density of a single Weyl neutrino: $\rho_{\nu_L} =(7 \pi^2/120)
(4/11)^{4/3} T_\gamma^4$, where $\rho_\gamma$ is the energy density of
photons (with temperature $T_\gamma$) and $\rho_{\rm R}$ is the total
energy density in relativistic particles~\cite{Steigman:1977kc}. A
selection of the most recent measurements of $N_{\rm eff}$ together
with the $1\sigma$ confidence intervals from various combinations of
models and data sets are shown in Fig.~\ref{fig:uno}. The data hint at
the presence of an excess $\Delta N$ above SM expectation of $N_{\rm
  eff} = 3.046$~\cite{Mangano:2005cc}.

One of the most striking results of the Planck spacecraft is that the
best-fit Hubble constant has the value $h= 0.674 \pm
0.012$~\cite{Ade:2013lta}.\footnote{We adopt the usual convention of
  writing the Hubble constant at the present day as $H_0 = 100 \
  h~{\rm km} \ {\rm s}^{-1} \ {\rm Mpc}^{-1}$. For $t= {\rm today}$,
  the various energy densities are expressed in units of the critical
  density $\rho_c$; {\it e.g.}, the matter density $\Omega_{\rm M}
  \equiv \rho_{\rm M}({\rm today})/\rho_c$.}  This result deviates by
more than 2$\sigma$ from the value obtained with the Hubble Space
Telescope, $h = 0.738 \pm 0.024$~\cite{Riess:2011yx}.  The impact of
the new $h$ determination is particularly complex in the investigation
of $N_{\rm eff}$. Combining CMB observations with data from baryon
acoustic oscillations (BAO) the Planck Collaboration reported $N_{\rm
  eff} = 3.30 \pm 0.27$. Adding the $H_0$ measurement to the CMB data
gives $N_{\rm eff} =3.62 \pm 0.25$ and relieves the tension between
the CMB data and $H_0$ at the expense of new neutrino-like physics (at
around the $2.3 \sigma$ level).  We have nothing further to add to
this discussion, but it seems worthwhile to analyze any other
independent set of data~\cite{Said:2013hta}.

 Several models have been proposed to explain $\Delta N$. To
 accommodate recent Planck observations in this brief 
 communication we re-examine two of them: models based on milli-weak
 interactions of right-handed partners of three Dirac
 neutrinos~\cite{Anchordoqui:2011nh}, and models in which the extra
 relativistic degrees of freedom are related to possible dark matter
 (DM) candidates produced via decay of heavy
 relics~\cite{Menestrina:2011mz}.\footnote{$\nu_R$'s have long been
   suspected to contribute to $N_{\rm eff}$~\cite{Ellis:1985fp}.} An
 interesting consequence of such a non-thermal DM scenario is that if
 the lifetime of the decaying particle is longer than about $10^3$
 seconds, the expansion history of the universe during the era of BBN
 will not have DM contributions to $N_{\rm eff}$.  Moreover, if there
 is a light DM partcile that annihilates to photons after the $\nu_L$
 have decoupled, the photons are heated beyond their usual heating
 from $e^\pm$ annihilation, reducing the late time ratio of neutrino
 and photon temperatures (and number densities), leading to $N_{\rm
   eff} < 3$~\cite{Ho:2012ug}. This opens the window for addition of
 one or more $\nu_R$ neutrino flavors and still be consistent with $N_{\rm
   eff} = 3$.

We begin by first establishing the contribution of right-handed
neutrinos to $N_{\rm eff}$, that is $\Delta N_\nu$ as a function of
the $\nu_R$ decoupling temperature.  A critical input for $\Delta
N_\nu$ is the relation between the r.d.o.f.  and the temperature of
the primordial plasma. This relation is complicated because the
temperature which is of interest for right-handed neutrino decoupling
from the heat bath may lie in the vicinity of the quark-hadron
cross-over transition.  To connect the temperature to an effective
number of r.d.o.f. we make use of some high statistics lattice
simulations of a QCD plasma in the hot phase, especially the behavior
of the entropy during the changeover~\cite{Bazavov:2009zn}.

\begin{figure}[!t]
  \centering
  \includegraphics[width=0.45\textwidth]{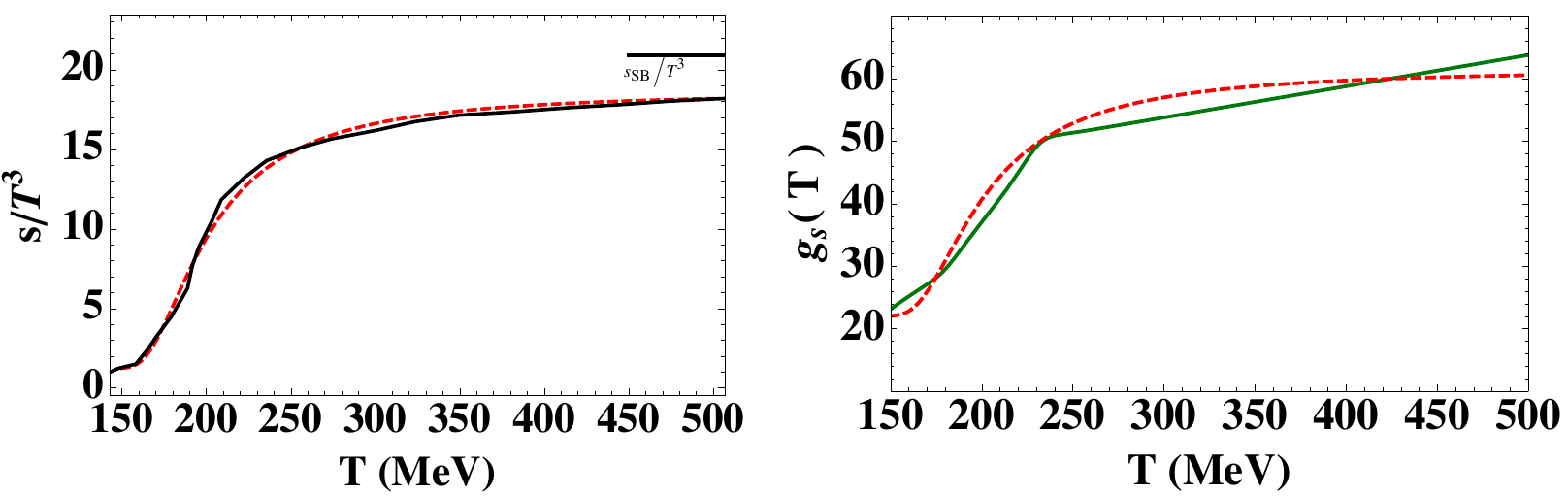}
  \caption{{\bf Left:} The parametrization of the entropy density
    given in Eq.~(\ref{soverT})  (dashed line) superposed on the
    result from high statistics lattice
    simulations~\cite{Bazavov:2009zn} (solid line). {\bf Right:}~Comparison of $g_s (T)$ obtained using Eq.~(\ref{gdet}) (dashed
    line) and the phenomenological estimate of~\cite{Laine:2006cp} (solid line).}
\label{fig:dos}
\end{figure}

Taking into account the isentropic heating of the rest of the plasma
between $T_{\nu_R}^{\rm dec}$ and $T_{\nu_L}^{\rm dec}$ decoupling
temperatures we obtain $ \Delta N_\nu = 3 [g_s(T_{\nu_L}^{\rm
  dec})/g_s(T_{\nu_R}^{\rm dec})] ^{4/3},$ where $g_s(T)$ is the
effective number of interacting (thermally coupled) r.d.o.f. at
temperature $T$; {\it e.g.}, $g_s(T_{\nu_L}^{\rm dec}) = 43/4$.  At
energies above the deconfinement transition towards the quark gluon
plasma, quarks and gluons are the relevant fields for the QCD sector,
such that the total number of SM r.d.o.f. is $g_s = 61.75$.  As the
universe cools down, the SM plasma transitions to a regime where
mesons and baryons are the pertinent degrees of freedom. Precisely,
the relevant hadrons present in this energy regime are pions and
charged kaons, such that $g_s = 19.25$~\cite{Brust:2013ova}. This
significant reduction in the degrees of freedom results from the rapid
annihilation or decay of any more massive hadrons which may have
formed during the transition. The quark-hadron crossover transition
therefore corresponds to a large redistribution of entropy into the
remaining degrees of freedom. Concretely, the effective number of
interacting r.d.o.f. in the plasma at temperature $T$ is given by $g_s
(T) \simeq r (T) \left(N_B+ \frac{7}{8} N_{\rm F} \right),$ with
$N_{\rm B} = 2$ for each real vector field and $N_{\rm F} = 2$ for
each spin-$\frac{1}{2}$ Weyl field. The coefficient $r (T)$ is unity
for leptons, two for photon contributions, and is the ratio $s(T )/s_{\rm
  SB}$ for the quark-gluon plasma. Here, $s(T)$ $(s_{\rm SB})$ is the
actual (ideal Stefan-Bolzmann) entropy shown in Fig~\ref{fig:dos}. The
entropy rise during the confinement-deconfinement changeover can be
parametrized, for $150~{\rm MeV} < T < 500~{\rm MeV}$, by
\begin{eqnarray}
\frac{s}{T^3} & \simeq & \frac{42.82}{\sqrt{392 \pi}} e^{-\frac{\left( T_{\rm MeV} - 151 \right)^2}{392}}  + \left( \frac{195.1}{T_{\rm MeV} - 134} \right)^2 \nonumber
\\
& \times & 18.62 \frac{e^{195.1/(T_{\rm MeV} -134)}} {\left[
    e^{195.1/(T_{\rm MeV} -134)} - 1\right]^2} \, .
\label{soverT}
\end{eqnarray}
For the same energy range, we obtain
\begin{equation}
g_s(T) \simeq 47.5 \ r(T) + 19.25 \, .
\label{gdet}
\end{equation}
In Fig.~\ref{fig:dos} we show $g_s(T)$ as given by (\ref{gdet}). Our
parametrization is in very good agreement with the phenomenological
estimate of~\cite{Laine:2006cp}.

\begin{figure}[!t]
  \centering
  \includegraphics[width=0.45\textwidth]{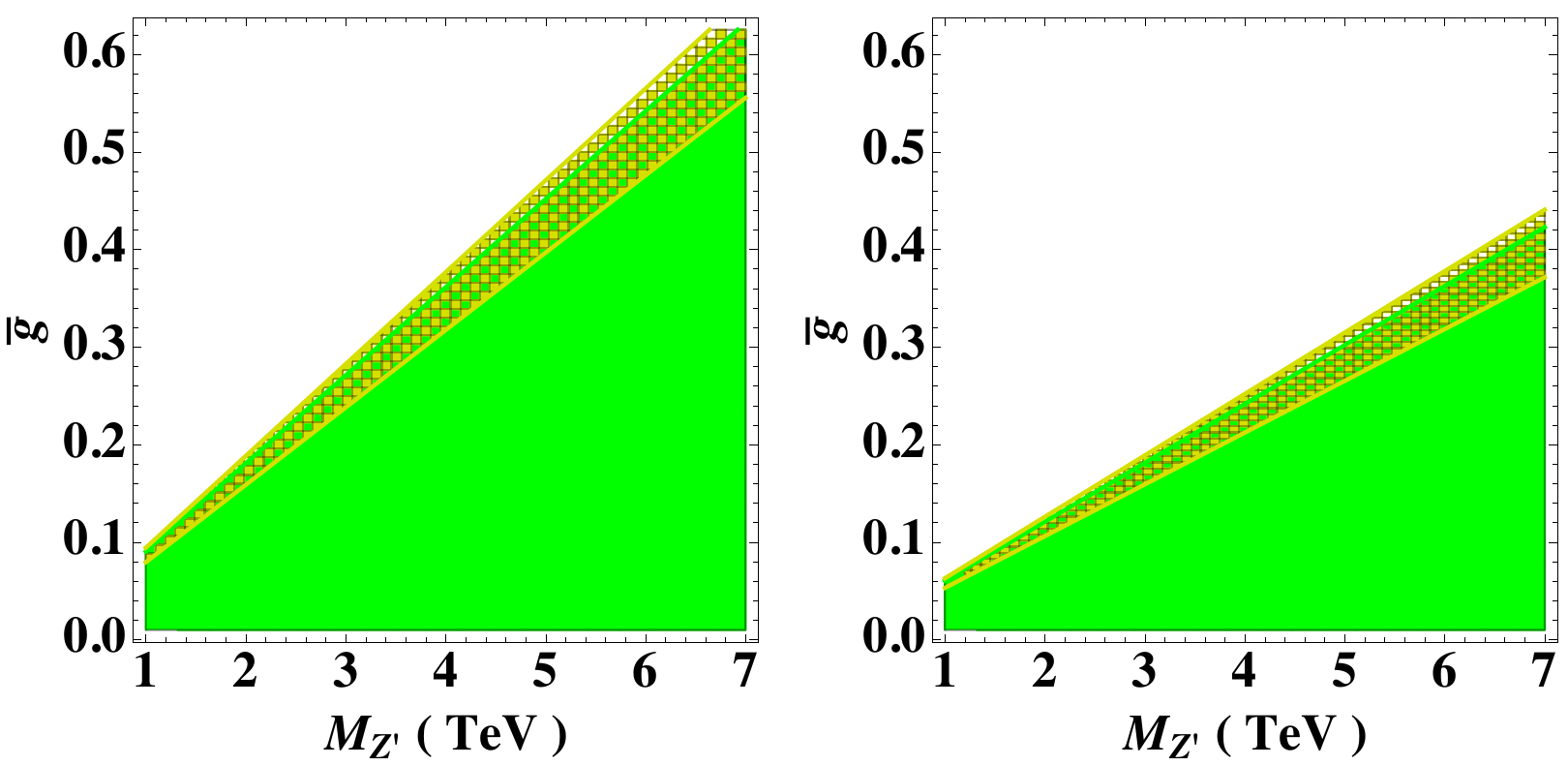}
  \caption{The (green) shaded area shows the $1\sigma$ region allowed
    from decoupling requirements to accommodate Planck + BAO, whereas
    the (yellow) cross-hatched area shows the region allowed from
    decoupling requirements to accommodate Planck + $H_0$. We have
    taken ${\cal K} =0.5$ (left) and ${\cal K} = 2.5$ (right).}
\label{fig:tres}
\end{figure}

If relativistic particles are present that have decoupled from the
photons, it is necessary to distinguish between two kinds of r.d.o.f.:
those associated with the total energy density $g_\rho$, and those
associated with the total entropy density $g_s$. Since the quark-gluon
energy density in the plasma has a similar $T$ dependence to that of
the entropy (see Fig. 7 in~\cite{Bazavov:2009zn}), we take $ g_\rho
(T) \simeq r (T) \left(N_B+ \frac{7}{8} N_{\rm F} \right)$.

The right-handed neutrino decouples  from the plasma when its mean
free path becomes greater than the Hubble radius at that time 
\begin{equation}
 \Gamma(T_{\nu_R}^{\rm dec}) = H(T_{\nu_R}^{\rm dec}) \, ,
\label{haim1}
\end{equation}
where
\begin{equation}
\Gamma (T_{\nu_R}^{\rm dec}) = {\cal K} \ \frac{1}{8} \ \left(\frac{\overline  g}{M_{Z'}}
\right)^4 \ (T_{\nu_R}^{\rm dec})^5 \ \sum_{i=1}^6 {\cal N}_i \,,
\label{Gamma}
\end{equation}
is the $\nu_R$ interaction rate, 
\begin{eqnarray}
H(T^{\rm dec}_{\nu_R}) & = & 1.66 \sqrt{g_\rho} \ (T_{\nu_R}^{\rm dec})^2
/M_{\rm Pl} \nonumber \\
& \simeq &
1.66 \sqrt{g_s (T^{\rm dec}_{\nu_L}) } \
\frac{(T^{\rm dec}_{\nu_R})^2}{M_{\rm
    Pl}} \ \left( \frac{3}{\Delta
  N_\nu} \right)^{3/8} \,,
\label{Hubble}
\end{eqnarray}
$ \overline g \equiv [(\sum_{i =1}^6 {\cal N}_i g_i^2 g_6^2) /
(\sum_{i =i}^6 {\cal N}_i )]^{1/4}$, ${\cal N}_i$ is the number of
chiral states, $g_i$ are the chiral couplings of the $Z'$ for the 6 relevant species (see below), and the
cosntant ${\cal K} = 0.5 \ (2.5)$ for annihilation (annihilation +
scattering)~\cite {Anchordoqui:2011nh}. In the second line of
(\ref{Hubble}) we set $g_s \simeq g_\rho$.

The physics of interest takes place in the quark gluon plasma itself
so that we will restrict ourselves to the following fermionic fields
in the visible sector, $ \left[ 3 u_R \right] + \left[ 3 d_R \right] +
\left[ 3s_R\right ] + \left[ 3 \nu_L + e_L + \mu_L \right] + \left[
  e_R + \mu_R \right]+ \left[ 3 u_L + 3d_L + 3s_L\right]+ \left[ 3
  \nu_R \right]$, and their contribution to $g_\rho$. This amounts to
28 Weyl fields, translating to 56 fermionic r.d.o.f.; $\sum_{i=1}^6
{\cal N}_i = 28$. To illustrate we calculate $\overline g$ for two
candidate models. The first is a set of variations on D-brane
constructions which do not have coupling constant unification, $0.29
\leq \overline g \leq 0.39$~\cite{Anchordoqui:2011eg}. The second are
$U(1)$ models which are embedded in a grand unified exceptional $E_6$
group, $ 0.09 \leq \overline g \leq 0.46$~\cite{Ellis:1985fp}.

Substituting (\ref{Gamma}) and
(\ref{Hubble}) into (\ref{haim1}) we obtain
\begin{displaymath}
\frac{\overline g}{M_{Z'}} = \left( \frac{3}{\Delta N_\nu}\right)^{3/32}
\left(\frac{13.28 \ \sqrt{g_s(T_{\nu_L}^{\rm dec})}}{M_{\rm Pl} \ {\cal
      K} \ (T_{\nu_R}^{\rm dec})^3 \  \sum_{i=1}^6 {\cal N}_i} \right)^{1/4} 
\end{displaymath}
and
\begin{equation}
 \Delta N_\nu 
 =  \left[\frac{5.39 \times 10^{-6}}{{\cal K} \sum_{i=1}^6 \mathcal{N}_i} \left(  \frac{M_{Z'}}{{\rm TeV}} \ \frac{1}{\bar{g}} \right)^4 \ \left(\frac{{\rm GeV}}{T_{\nu_R}}\right)^3 \right]^{8/3}  \, .
\label{hitichi}
\end{equation}
In Fig.~\ref{fig:tres} we show the region of the parameter space
allowed from decoupling requirements to accommodate contributions of
$\Delta N_\nu$ within the $1\sigma$ region of Planck data. It is
important to stress that the LHC experimental limits on $M_{Z'}$ for
null signals for enhancements in dilepton searches entail $M_{Z'} >
2.3~{\rm TeV}$ at the 95\%~CL~\cite{Chatrchyan:2012oaa}, whereas
limits from dijet final states entail $M_{Z'} > 4~{\rm TeV}$
at 95\%~CL~\cite{Chatrchyan:2013qha}.

Next, in line with our stated plan, we derive quantitative bounds on
the fraction of non-thermally produced DM particles, which are
relativistic at the CMB epoch. The assumption herein is that of the
total DM density around today, $\Omega_{\rm DM}$, a small fraction,
$f = \Omega_X/\Omega_{\rm DM}$, is of particles
of type $X$, produced via decay of a heavy relic $X'$ with mass $M'$
and lifetime $\tau$: $X' \rightarrow X + \gamma$.\footnote{We do not at present
consider the more complicated scenario in which high energy neutrinos
are among the decay products~\cite{GonzalezGarcia:2012yq}.} At any time after the decay of $X'$
the total DM energy density is found to be
\begin{equation}
\rho_{\rm DM} (t) = \frac{M \ n_X ({\rm today})}{a^3(t)} \ \gamma (t) +
(1-f) \ \rho_c \ \frac{\Omega_{\rm DM}}{a^3(t)} \,,
\end{equation}
where $M$ is the mass of the $X$ particle, $n_X(t)$ its number density,
and $a(t)$ is the expansion scale factor normalized by $a({\rm today}) = 1$.
The scale factor dependence on the Lorentz boost is given by
\begin{eqnarray}
\gamma(t)   & = & \sqrt{ 1 + \left[a (\tau)/a(t) \right]^2 \left[ \gamma^2 (\tau) -
  1\right]}  \nonumber \\
& \approx & 1 + \frac{1}{2} \left[a
  (\tau)/a(t) \right]^2 \left[\gamma^2 (\tau) -1 \right] - \frac{1}{8} \left[
a (\tau)/a(t) \right]^4 \nonumber \\
& \times & \left[ \gamma^2 (\tau) -1 \right]^2 \ + \cdots \, , 
\label{expansion}
\end{eqnarray}
where
\begin{equation}
\gamma (\tau) = \frac{M'}{2 M}+\frac{M}{2M'} \, .
\end{equation}
At the present day the $X$ particles are non-relativistic, {\it i.e.}
$\Omega_X ({\rm today}) =  M \ n_X ({\rm today})/\rho_c$.  To obtain
such a non-relativistic limit we demand the magnitude of the second
term in the expansion of (\ref{expansion}) to be greater than the
third term, which results in $[a (\tau)/a(t)]^2 [\gamma^2 (\tau) -1] <
4$. Contrariwise, by this criteria the particle $X$ is relativistic if
$\gamma (t) > \sqrt{5}$.

\begin{figure}[ht]
  \centering
  \includegraphics[width=0.45\textwidth]{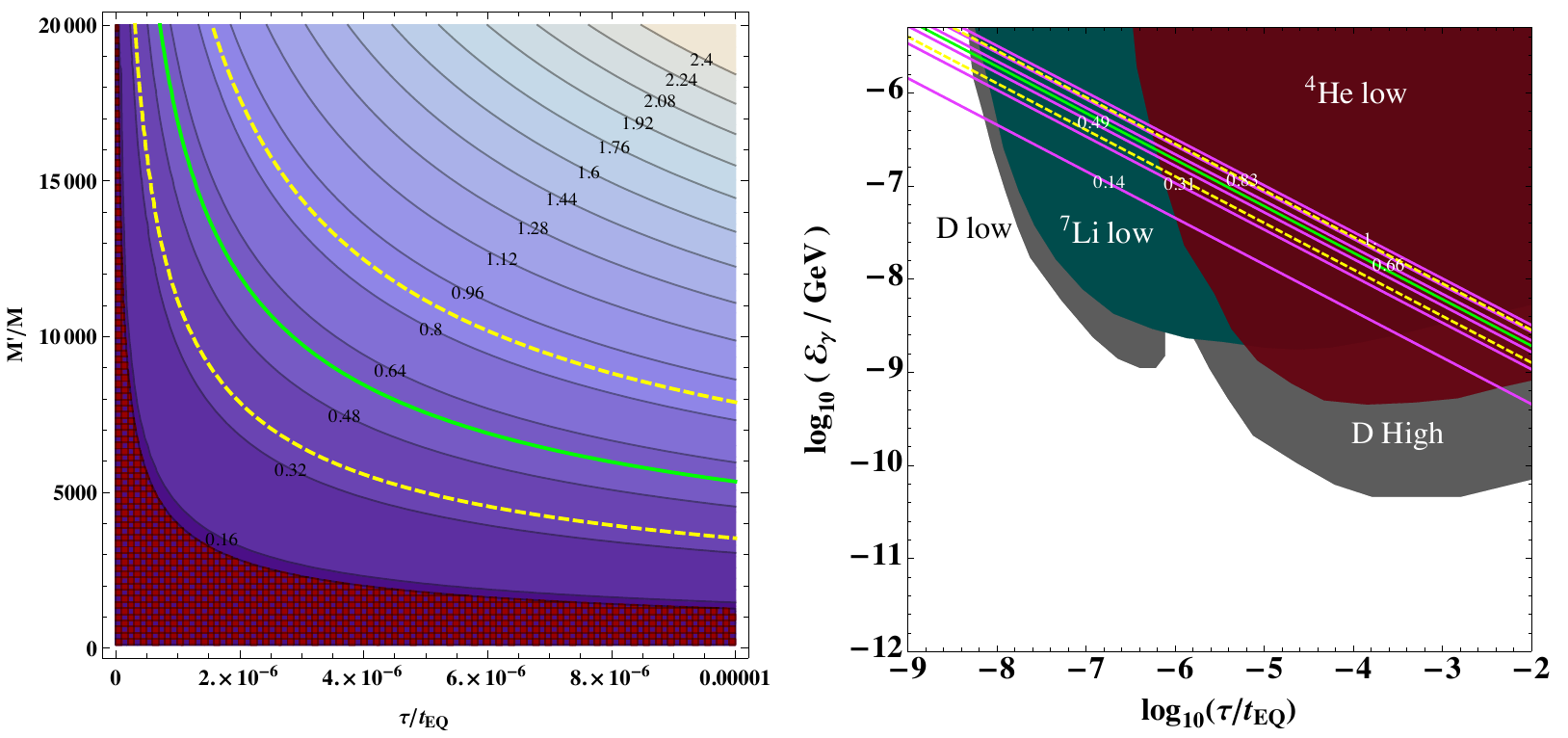}
  \caption{{\bf Left:} Contours of constant $\Delta N_X$ in the $R -
    \tau/t_{\rm EQ}$ plane, with $f = 0.01$, for the case in which
    $\Delta N_\nu = 0$. The (green) solid line indicates the upper
    limit on $\Delta N$ from Planck + BAO. The band between the dashed
    (yellow) lines corresponds to the allowed $\Delta N$ from Planck +
    $H_0$. The crosshatched area pertains to the region of the
    parameter space for which the $X$ particles are non-relativistic
    at the CMB epoch, and should be therefore not taken under
    consideration. {\bf Right:} The shaded regions show BBN
    constraints on the decay $X' \to X +
    \gamma$~\cite{Cyburt:2002uv}. The various contributions to $\Delta
    N_X$ fall on straight lines.}
\label{fig:cuatro}
\end{figure}

To obtain the $X$ component of $N_{\rm eff}$ we assume that $X$
particles decouple from the plasma prior to $\nu_L$ decoupling,
conserving the ratio $T_\gamma/T_{\nu_L}$ from SM cosmology,
\begin{eqnarray}
\Delta N_X & = & \frac{8}{7} \left(\frac{11}{4} \right)^{4/3} \frac{\rho_X (t_{\rm EQ}) }{\rho_\gamma (t_{\rm EQ})} \nonumber \\
& = & \frac{8}{7} \ \left( \frac{11}{4} \right)^{4/3} \ \frac{\Omega_{\rm DM}}{\Omega_\gamma} \ a(t_{\rm EQ}) \ \gamma(t_{\rm EQ}) \ f \, ,
\label{deltanX1}
\end{eqnarray}
where $\rho_\gamma(\teq) = \rho_c \Omega_\gamma/a^4(\teq)$. The
radiation-matter equality time is defined by the condition $\rho_{\rm
  R} = \rho_{\rm M}$, which implies
 \begin{equation}
 \frac{1}{a(t_{\rm EQ})} \frac{\Omega_\gamma}{\Omega_{\rm M}} \left[1 + \frac{7}{8}\left(\frac{4}{11}\right)^{4/3} \left( 3 + \Delta N \right) \right] = 1 \,,
\label{teq}
 \end{equation} 
 with $\Delta N = \Delta N_\nu + \Delta N_X$. By expressing the scale factor in a piece-wise form,
\begin{displaymath}
a(t)  = \left\{ \begin{array}{ll} \Big( \frac{3}{2} \ H_0 \ t \Big)^{2/3} & \ \ \ {\rm if} \ t > t_{\rm EQ}  \\
 \left( \frac{3}{2} \ H_0 \ t_{\rm EQ}^{1/4} \right)^{2/3} \ t^{1/2}  & \ \ \ {\rm if} \  t_{\rm EQ} > t > 1~{\rm s} \end{array} \right. , \nonumber
\end{displaymath}
from (\ref{teq}) we obtain
\begin{displaymath}
 h t_{\rm EQ}= \frac{ {\rm s} \, {\rm Mpc}}{150~{\rm km}}  \left\{ \frac{\Omega_{\gamma}}{\Omega_{\rm M}} \left[1 + \frac{7}{8}\left(\frac{4}{11}\right)^{4/3} \left( 3 + \Delta N \right) \right ]\right\}^{3/2}. 
\end{displaymath}
Note that the combination $h t_{\rm EQ}$ is independent of $h$. 

At this point it is worth exploring the quanity $\gamma(t_{\rm EQ})$ and its dependence on $h$,
\begin{eqnarray}
\gamma(t_{\rm EQ}) & = & \sqrt{1 + \left[a(\tau)/a(t_{\rm
        EQ})\right]^2 \ \left( R^2 - 1\right)^2/(4 R^2)} \nonumber \\
& = & \sqrt{1 + \left(\tau/t_{\rm EQ}\right) \
 \left( R^2 - 1\right)^2/(4 R^2)} \, ,
\label{treintaicuatro}
\end{eqnarray} 
where $R = M'/M$ and we have assumed that $X'$ decays before the time
of matter radiation equality. Interestingly, if time is scaled to
$t_{\rm EQ}$ the analysis is $h$ independent. This is important
because of the recent tension between astrophysical data and
cosmological observations. The systematics on $h$ will seriously
influence the dependence on $N_{\rm eff}$. Putting all this together
(\ref{deltanX1}) can be rewritten as
\begin{equation}
 \Delta N_X   = \frac{\Omega_{\rm DM}}{\Omega_{\rm M} } \ {\cal A} \  \left[
\frac{8}{7} \left( \frac{11}{4}
 \right)^{4/3} +   3 + \Delta N \right] f.
 \label{trentaisiete}
 \end{equation}  
where
\begin{equation}
{\cal A} = \sqrt{1 + \frac{\tau}{t_{\rm EQ}} \frac{\left( R^2 - 1\right)^2}{4 R^2}} \, .
\end{equation} 
Note that (\ref{trentaisiete}) scales with the ratio $\Omega_{\rm
  DM}/\Omega_{\rm M} \simeq 0.84$, which does not depend on the Hubble
parameter. Finally, solving (\ref{trentaisiete}) for $\Delta N_X$ gives
\begin{equation}
 \Delta N_X = \left(7.4 + \Delta N_\nu  \right) \ {\cal A}  \ \left(\frac{\Omega_{\rm M}}{\Omega_{\rm DM} \ f} -  {\cal A} \right)^{-1} .
 \label{trentaiocho}
 \end{equation}
In Fig.~\ref{fig:cuatro} we show contours of constant $\Delta N_X$ in
 the $R - \tau/t_{\rm EQ}$ plane, for the case in which $\Delta
 N_\nu = 0$. As expected, to produce a given $\Delta N_X$
 contribution, the required ratio of masses diminishes with increasing
 lifetime.

 We now verify that $X' \to X + \gamma$ does not drastically alter any
 of the light elemental abundances synthesized during
 BBN. Following~\cite{Cyburt:2002uv}, we assume the photons injected
 into the plasma rapidly redistribute their energy through scattering
 off background photons and through inverse Compton scattering. As a
 consequence, the constraints from BBN are (almost) independent of the
 initial energy distribution of the injected photons, and are only
 sensitive to the total energy released in the decay process. In the
 spirit of~\cite{Feng:2003uy}, we conveniently write the
 electromagnetic energy release as $\varepsilon_\gamma \equiv E_\gamma
 Y_{X'}$, where $E_{\gamma} = ({M'}^2 - M^2)/(2 M')$ is the initial
 electromagnetic energy release in each $X'$ decay and $Y_{X'} \equiv
 n_{X'}/n_\gamma^{\rm BG}$ is the number density of $X'$ before the
 decay, normalized to the number density of background photons
 $n_\gamma^{\rm BG} = 2 \ \zeta(2) \ T_\gamma^3/ \pi^2$.  For
 $Y_{X'}$, each $X'$ decay produces one $X$, and so the $X'$ abundance
 may be expressed in terms of the present $X$ abundance through
\begin{eqnarray}
Y_{X'} (\tau)  \! \! \!\! & = & \! \! \!\! Y_{X,  \tau} = Y_{X, {\rm today}} = \frac{\Omega_{X} \rho_c}{M \ n_{\gamma}^{\rm BG} ({\rm today})} \nonumber \\
    & \simeq & \! \! \! \! 2.26 \times 10^{-14} \ \frac{\rm
        TeV}{M} \ \frac{\Omega_{\rm DM} h^2}{0.1199} \
      \frac{f}{0.01} \ ,
\label{Yx}
\end{eqnarray}
yielding
\begin{displaymath}
\varepsilon_\gamma  =  1.13 \times 10^{-11} \ \
  \frac{\Omega_{\rm DM} h^2}{0.1199} \ \  \frac{f}{0.01} \ \left( \frac{M'}{M} - \frac{M}{M'} \right)~{\rm GeV} \, .
\end{displaymath}
The thorough analysis of electromagnetic cascades reported
in~\cite{Cyburt:2002uv} reveals that the shaded regions of
Fig.~\ref{fig:cuatro} are ruled out by considerations of light
elemental abundances produced during BBN. The various regions are
disfavored by the following conservative criteria: {\it (i)}~D/H $<
10^{-4.9}$ (low); {\it (ii)}~D/H $> 10^{-4.3}$ (high);  {\it (iii)}~$^7$Li/H $< 10^{-10.05}$; {\it
  (iv)}~primordial $^4$He abundance $< 0.227$. The straight lines represent several
combinations of $R$ and $\tau/t_{\rm EQ}$ producing the $\Delta N_X$
indicated in the labels.  All straight lines intersect the BBN bounds
at about ${\rm log}_{10}(\tau/t_{EQ}) = -8.2$. The constraints from BBN are weak for early
decays because at early times the universe is hot and thus the $X'$
secondary photon spectrum is rapidly thermalized, leaving just a few
high-energy photons that cannot alter the light elemental
abundances. However, for $\tau/t_{\rm EQ} > 10^{-8.2}$ BBN excludes
most of the relevant parameter space.

\begin{figure}[!t]
  \centering
  \includegraphics[width=0.45\textwidth]{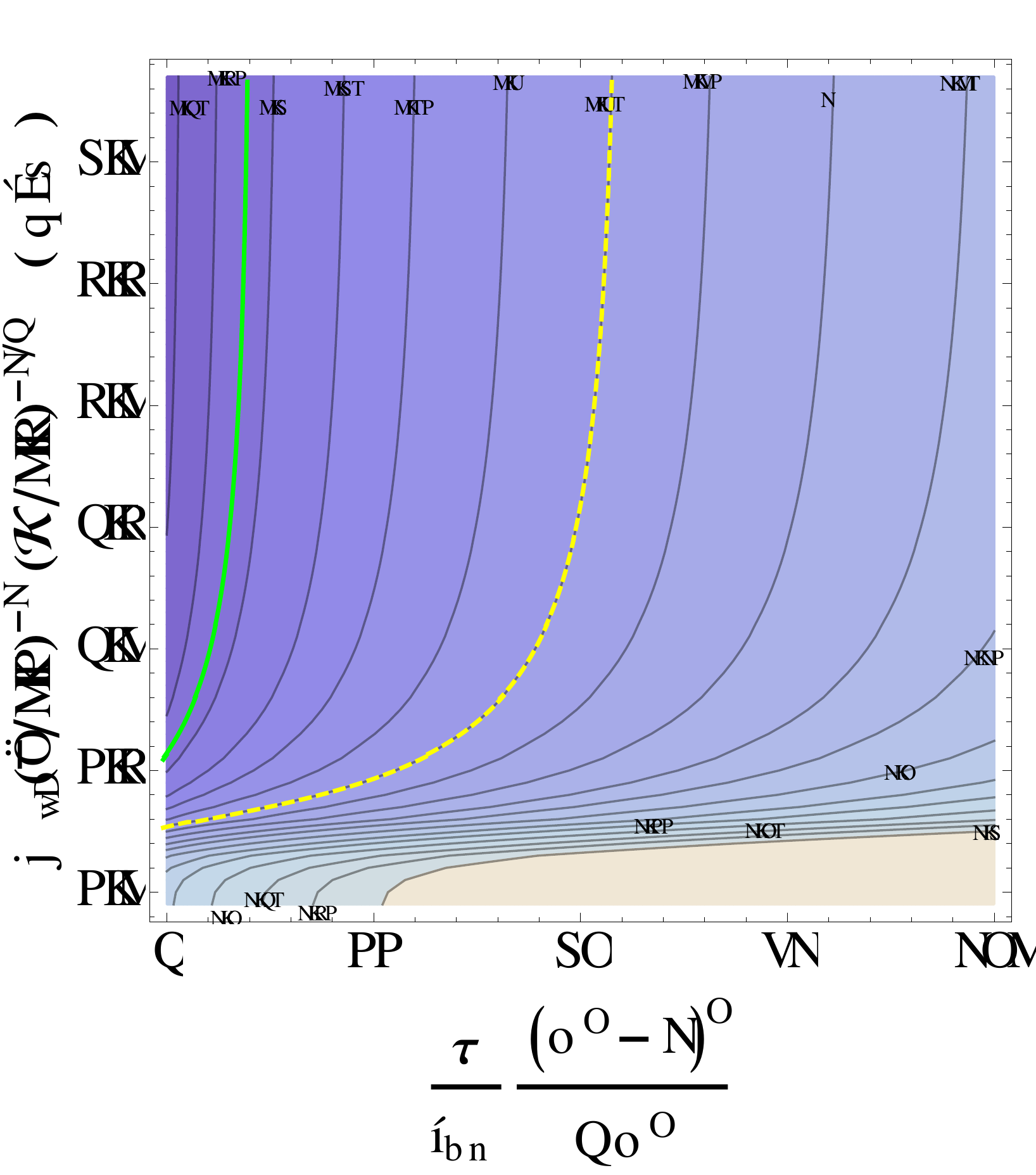}
  \caption{Contours of constant $\Delta N$ in the $M_{Z'}/ ({\cal K}\overline
    g)$ {\it vs.} $(\tau/t_{\rm EQ}) (R^2-1)^2/(2R)^2$ plane, with $f=
    0.01$. The solid (dashed) line indicates the $1\sigma$ upper limit
    on $\Delta N$ from Planck + BAO (Planck +
    $H_0$).}
  \label{fig:cinco}
 \end{figure}

 Finally, we combine (\ref{hitichi}) and (\ref{trentaiocho}) to
 account for both $\Delta N_\nu$ and $\Delta N_X$ components. In
 Fig.~\ref{fig:cinco} we show contours of constant $\Delta N$ in the
 $M_{Z'}/ (K\overline g)$ {\it vs.} $(\tau/t_{\rm EQ})
 (R^2-1)^2/(2R)^2$ plane. Interestingly, despite the very restrict
 limit from Planck + BAO data, there is room within the $1\sigma$
 allowed $N_{\rm eff}$ region to accommodate contributions from both
 non-thermal DM and right-handed neutrinos which interact with all
 fermions via a $Z'$ gauge boson that is within the LHC discovery
 reach.  Note that for (say) $\Delta N_X \approx 0.3$, the particular
 range of DM parameters includes ($\tau/t_{\rm EQ} \approx 10^{-8.2}$,
 $R \approx 6 \times 10^4$), which is not excluded by BBN
 considerations.

 \vspace*{0.5cm} \footnotesize{{\bf Acknowledgment:}{ LAA and VB are
     supported by NSF CAREER PHY-1053663. HG is supported by NSF
     PHY-0757959.


\begin{thebibliography}{}


\bibitem{Steigman:1977kc} 
  G.~Steigman, D.N.~Schramm and J.E.~Gunn,
  Phys.\ Lett.\ B {\bf 66}, 202 (1977).






\bibitem{Mangano:2005cc} 
  G.~Mangano, G.~Miele, S.~Pastor, T.~Pinto, O.~Pisanti and P.D.~Serpico,
  Nucl.\ Phys.\ B {\bf 729}, 221 (2005).


\bibitem{Ade:2013lta} 
  P.A.R.~Ade {\it et al.}  [Planck Collaboration],
  arXiv:1303.5076.

\bibitem{Riess:2011yx} 
  A.~G.~Riess {\it et al.},
  Astrophys.\ J.\  {\bf 730}, 119 (2011)  [Erratum ibid.\  {\bf 732}, 129 (2011)].

\bibitem{Said:2013hta} 
  N. Said, E. Di Valentino and M. Gerbino,
  arXiv:1304.6217.


\bibitem{Anchordoqui:2011nh} 
  L.A. Anchordoqui and H. Goldberg,
  Phys.\ Rev.\ Lett.\  {\bf 108}, 081805 (2012);
  A. Solaguren-Beascoa and M.C. Gonzalez-Garcia,
  Phys.\ Lett.\ B {\bf 719}, 121 (2013);
  L.A. Anchordoqui, H. Goldberg and G. Steigman,
  Phys.\ Lett.\ B {\bf 718}, 1162 (2013).








\bibitem{Menestrina:2011mz} 
  J.L. Menestrina and R.J. Scherrer,
  Phys.\ Rev.\ D {\bf 85}, 047301 (2012);
  D. Hooper, F.S. Queiroz and N.Y. Gnedin,
  Phys.\ Rev.\ D {\bf 85}, 063513 (2012);
  C. Kelso, S. Profumo and F.S. Queiroz,
  arXiv:1304.5243.

\bibitem{Ho:2012ug} 
  C.M. Ho and R.J. Scherrer,
  Phys.\ Rev.\ D {\bf 87}, 023505 (2013);
  G.~Steigman,
  arXiv:1303.0049. 


\bibitem{Ellis:1985fp} 
  J.R. Ellis, K. Enqvist, D.V. Nanopoulos and S. Sarkar,
  Phys.\ Lett.\ B {\bf 167}, 457 (1986);
  M.C. Gonzalez-Garcia and J.W.F. Valle,
  Phys.\ Lett.\ B {\bf 240}, 163 (1990);
  J.L. Lopez and D.V. Nanopoulos,
  Phys.\ Lett.\ B {\bf 241}, 392 (1990);
  V. Barger, P. Langacker and H.S. Lee,
  Phys.\ Rev.\  D {\bf 67}, 075009 (2003).


\bibitem{Bazavov:2009zn} 
  A.~Bazavov  {\it et al.},
  Phys.\ Rev.\ D {\bf 80}, 014504 (2009).



\bibitem{Brust:2013ova} 
  C. Brust, D.E. Kaplan and M.T. Walters,
  arXiv:1303.5379. 





\bibitem{Laine:2006cp} 
  M. Laine and Y. Schroder,
  Phys.\ Rev.\ D {\bf 73}, 085009 (2006);
  G. Steigman, B. Dasgupta and J.F. Beacom,
  Phys.\ Rev.\ D {\bf 86}, 023506 (2012).


\bibitem{Anchordoqui:2011eg}
  L.A. Anchordoqui, I. Antoniadis, H. Goldberg, X. Huang, D. Lust and T.R. Taylor,
Phys.\ Rev.\  D {\bf 85},  086003 (2012); 
  L.A. Anchordoqui, I. Antoniadis, H. Goldberg, X. Huang, D. Lust, T.R. Taylor and B. Vlcek,
  Phys.\ Rev.\ D {\bf 86}, 066004 (2012).



\bibitem{Chatrchyan:2012oaa} 
  S. Chatrchyan {\it et al.}  [CMS Collaboration],
  Phys.\ Lett.\ B {\bf 720}, 63 (2013);
  G. Aad {\it et al.}  [ATLAS Collaboration],
  JHEP {\bf 1211}, 138 (2012).


\bibitem{Chatrchyan:2013qha} 
  S. Chatrchyan {\it et al.}  [CMS Collaboration],
  arXiv:1302.4794;
  G. Aad {\it et al.}  [ATLAS Collaboration],
  JHEP {\bf 1301}, 029 (2013).




\bibitem{GonzalezGarcia:2012yq} 
  M.C. Gonzalez-Garcia, V. Niro and J. Salvado,
  arXiv:1212.1472.
   
\bibitem{Cyburt:2002uv} 
  R.H. Cyburt, J.R. Ellis, B.D. Fields and K.A. Olive,
  Phys.\ Rev.\ D {\bf 67}, 103521 (2003).




\bibitem{Feng:2003uy} 
  J.L. Feng, A. Rajaraman and F. Takayama,
  Phys.\ Rev.\ D {\bf 68}, 063504 (2003).





\end{thebibliography}
\end{document}